\documentclass[12pt,3p,twocolumn]{elsarticle}
\usepackage{graphicx,subfigure}
\usepackage{amsmath}
\def\degree{$^\circ$}
\title{Atmospheric effects on Quaternary polarization encoding for free space communication, laboratory simulation}
\author{Ram Soorat, Ashok Vudayagiri \\ School of Physics, University of Hyderabad, Hyderabad 500046}
\date{}
\begin{document}

\begin{abstract}
We have simulated atmospheric effects such as fog and smoke in laboratory environment to simulate depolarisation due to atmospheric effects during a free space optical communication. This has been used to study noise in  two components of quaternary encoding for polarization shift keying. Individual components of a Quaternary encoding, such as vertical and horizontal as well as 45$^\circ$ and 135$^\circ$, are tested separately and indicates that the depolarization effects are different for these two situation. However, due to a differential method used to extract information bits, the protocol shows extremely low bit error rates. The information obtained is useful during deployment of a fully functional Quaternary encoded PolSK scheme in free space. 
\end{abstract}

\maketitle

\section{Introduction}
Two-binary, one-quaternary (2B1Q) encoding is a method of  mapping a pair of bits (DiBIT) to single encoding of a 4-level scheme. Such methods of encoding multiple bits to each level of the encoding scheme allows increase in transmission density, although most of the times requires hardware with better precision. Several schemes for such multiple encoding have been proposed and studied earlier \cite{bhattacharya,tanay_opt_comm,tanay_j_opt} and their robustness have been discussed. Some of them are optical methods involving light, wherein different polarization or phase states of light are mapped to DiBITs \cite{tanay_opt_comm,tanay_j_opt}. However, when such methods are operated in free space communication protocols, concern has to be taken about the fact that atmospheric phenomena such as fog and smoke cause multiple scattering of the light, leading to a depolarization \cite{dadrasnia,smyth,zabidi,andrews,majumdar}. In additon, as we show in this communication, the depolarization behaviour is different for different encodings, adding additional difficulty, which is peculiar to M-ary encodings.  But our method of differential measurement, as shown earlier \cite{soorat}, will allow us to extract information with a near zero bit error rate despite a significant depolarization. 

The paper is organized as follows. We first briefly describe our experimental setup, which is same as the one used in \cite{soorat}. We then present the results for errors due to depolarization, first for vertical and horizontal encoding and then independently for 45$^\circ$ and 135\degree. The present study is not a complete and proper Quaternary scheme, but instead a depolarization study of individual components. But this will help us understand whether or not the atmospheric effects such as fog and smoke affect these states differently, and if so, whether that information can be incorporated into the communication scheme for better detection. 

We also explain the differential method of measurement based on State of Polarization, which allows us a higher tolerance for depolarization. Finally we show an analysis of this technique in terms of the Muller matrix method, which incorporates the depolarization effects. This allows us to represent the depolarized light as a Stokes Vector with partial polarization. The mathematical analysis for the differential method of measurement shows that the final measure is a value whose sign can be used to identify and map to the information bit.

\section{Experiment} 
\begin{figure}[h]
\subfigure[]{\includegraphics[scale=0.25]{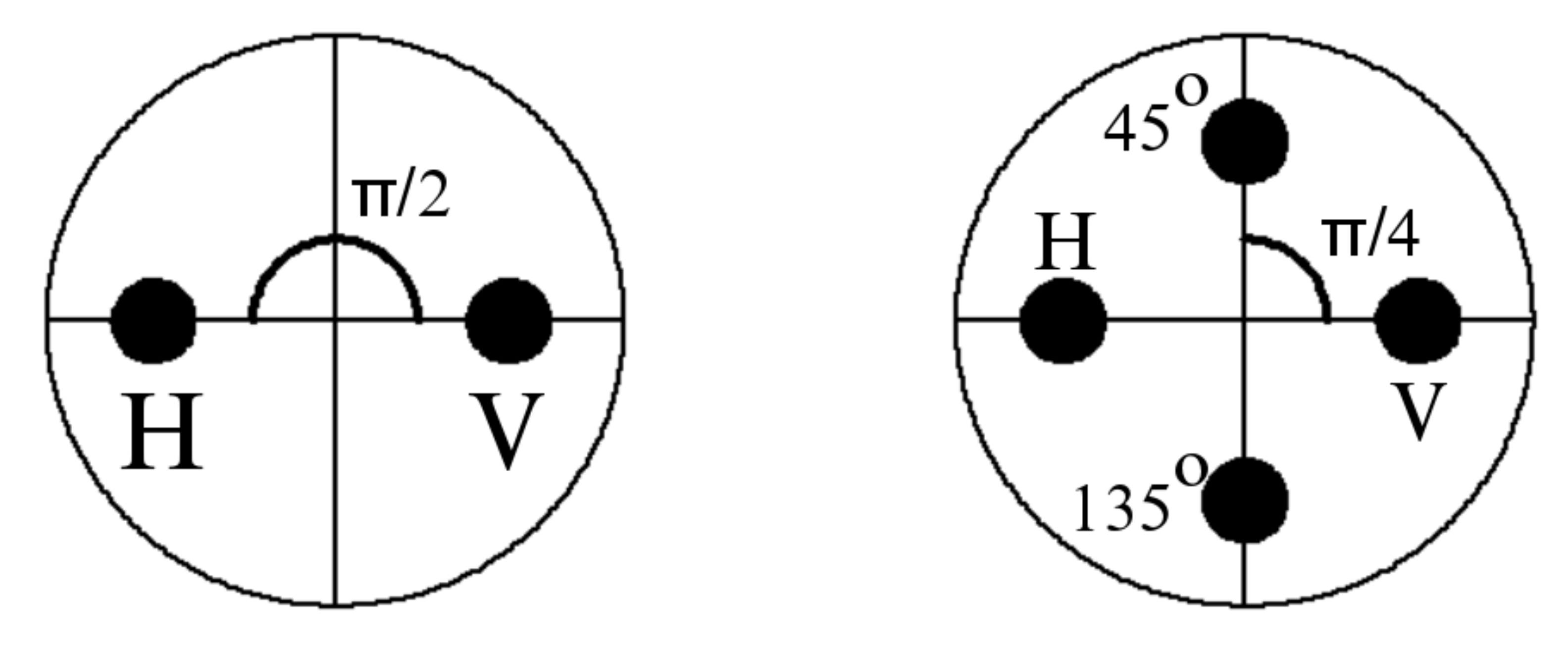}\label{constellation}}
\subfigure[]{\includegraphics[scale=0.25]{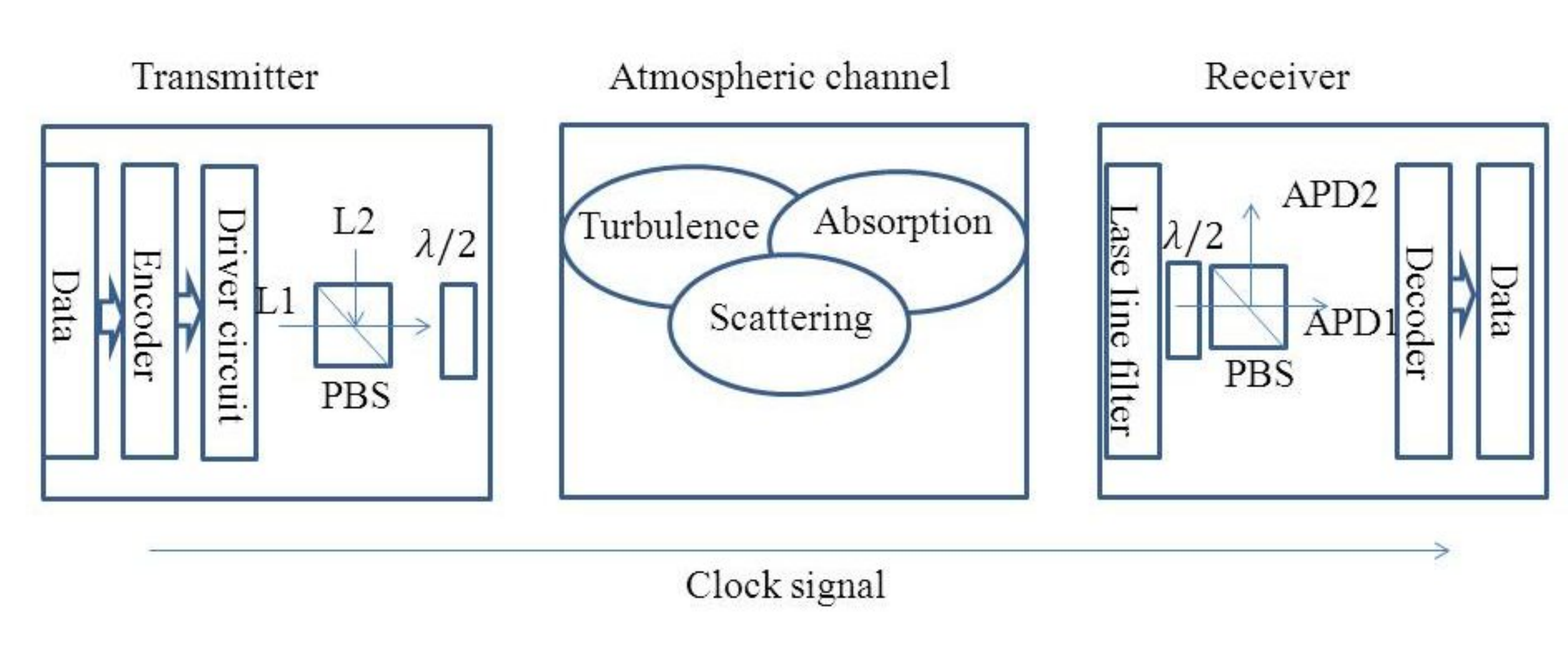}\label{setup}}
\caption{(a) Constellation diagrams for Binary (left) and Quaternary encoding (right) (b) Schematic of the experimental setup. See text for details.}
\label{setup}
\end{figure}

We consider a specific Quaternary encoding consisting of two pairs of mutually orthogonal polarizations - viz Horizontal and Vertical, as well as 45\degree and 135\degree. These can be represented on a standard constellation diagram as shown in figure \ref{constellation}. On the left is only the Binary encoding using only Horizontal and Vertical polarizations, while the right side shows the Quaternary encoding. Angle between them. When finally deployed, we should be able to use this as an encoding for pair of bits as 

\begin{table}[!h]
\begin{tabular}{r|l}
{\bf bit pairs} & {\bf Polarization} \\ \hline
00 & vertical \\
01 & 45\degree \\
10 & 135 \degree \\
11 & horizontal 
\end{tabular} 
\caption{Quaternary encoding of bit pairs to polarization}
\label{encoding}
\end{table}

	The experimental setup is similar to the one explained in our earlier paper \cite{soorat},  briefly explained here for sake of completeness. It consists of two lasers L1 and L2, both VCSEL's operating at 780nm. The choice of the wavelength is due to the fact that atmospheric attenuation has a clear window in this region \cite{gisin}. Light from lasers L1 and L2 is split into two polarized components by the polarizing beamsplitter PBS and vertical part of light from L1  and horizontal part of light from L2 goes into the communication channel. A LabVIEW software controls the lasers through the driver circuit, such that bit `0' results in a pulse from L1 and bit `1' results in a pulse from L2. The Halfwave plate, indicated by $\lambda/2$ in figure, is used to rotate the polarization whenever required. This is used to choose between vertical/horizontal scheme or 45\degree/135\degree scheme as required. 

	A glass chamber placed in the communication channel is filled with fog or smoke to simulate the atmospheric conditions. 
Smoke is created in the chamber by burning household incense powder. Fog is achieved by sprinkling water onto a a sample dry ice kept within. 	Both smoke and fog cause depolarization of the light due to  multiple scattering by smoke particles or water droplets. This results in a loss of information. 

The amount of smoke or fog is quantified by measuring the attenuation of the laser after passing through and characterized by optical density in dB, as given by \footnote{An alternative definition of Optical density is in terms of Beer - Lambert law, as $\alpha=\ln I/I_0$, which serves the function equally well. In this manuscript, we continue to use the definition as in attenuation} 	

$${\rm OD(dB)}=10\log\left(\frac{\rm Transmitted~Intensity}{\rm Incident~intensity}\right).$$

The receiver consists of another polarizing beamsplitter, labelled PBS2 and two Avalanche Photodiodes (labeled APD1 and APD2, both PCD Mini 0020 module from SenSL with 20 $\mu$m sensor and peltier cooler). The TTL pulses produced by the APD's are recorded via a DAQ card and counted using LabVIEW program. The counters are synchronized to a stream of clock pulses from the transmitter. To characterize the entire system for all possible polarization schemes, a random sequence of ones and zeros were generated by the computer and transmitted. The transmitted and received data are compared to obtain true Bit Error Rate (BER). These bit error rates are compared with the theoretical estimate of the BER for normal OOK scheme.

\begin{figure}[!h]
\centerline{\includegraphics[scale=0.5]{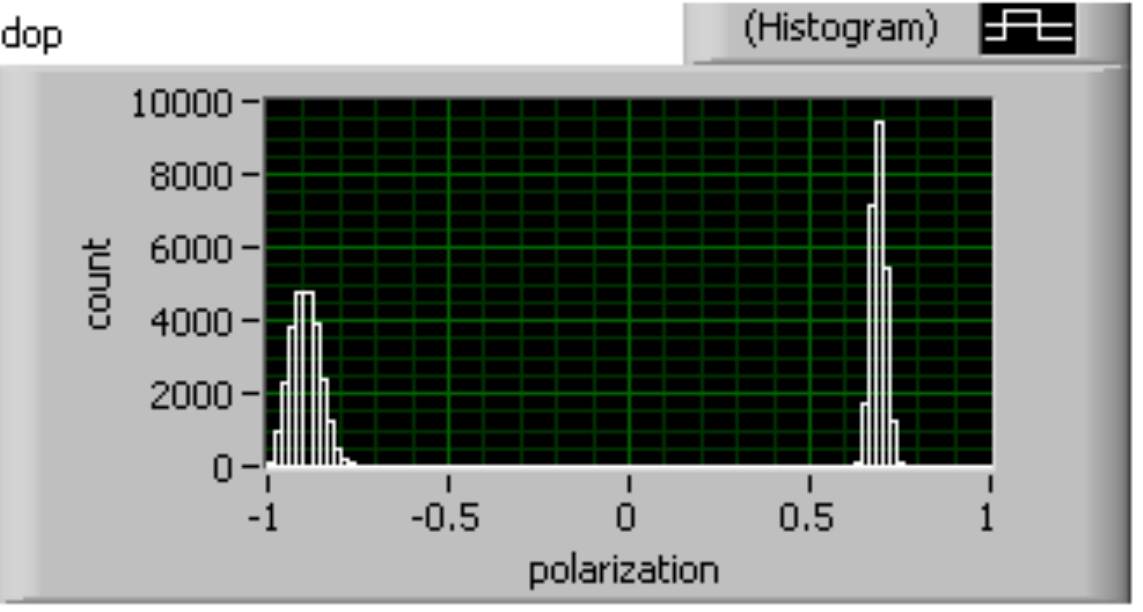}}
\caption{Statistical Distribution of SOP values}
\label{sop_histogram}
\end{figure}

The LabVIEW program computes State of Polarization, which is defined by us as 

\begin{equation}
S=\frac{\rm APD2 - APD1}{\rm APD2 + APD1} \label{sop}
\end{equation}

where APD1 and APD2 indicate the counts of photons on the two detectors. In the ideal case with no depolarization, a horizontally polarized light will all reach APD1, resulting in an SOP of $-1$, and a vertically polarized light will all reach APD2, resulting in an SOP of $+1$. However, depolarization causes some `leakage' to the other APD, resulting in an SOP value lesser than the range $\pm 1$. Since multiple scattering are random events, each pulse of similar polarization results in different values of SOP. A distribution of such several SOP values obtained from a single run of about 10000 pulses is shown in figure \ref{sop_histogram}. It  clearly shows the effect of depolarization, wherein the mean values are shifted inward from $\pm 1$. But since the two distributions are not overlapping each other, an unambiguous bit designation can still be performed, which shows a higher tolerance for noise, as shown in our earlier work \cite{soorat}. We have also shown in the earlier work that the measure of SOP obtained, is actually due to due to the ballistic and snake photons \cite{alfano, hema, soorat} whereas those that undergo significant depolarization cancel out each other. In this communication, we test this method with other polarizations, which will form components of a Quaternary encoding.

\section{Results \& Discussions}

\begin{figure}[!h]
\subfigure[]{\includegraphics[scale=0.2]{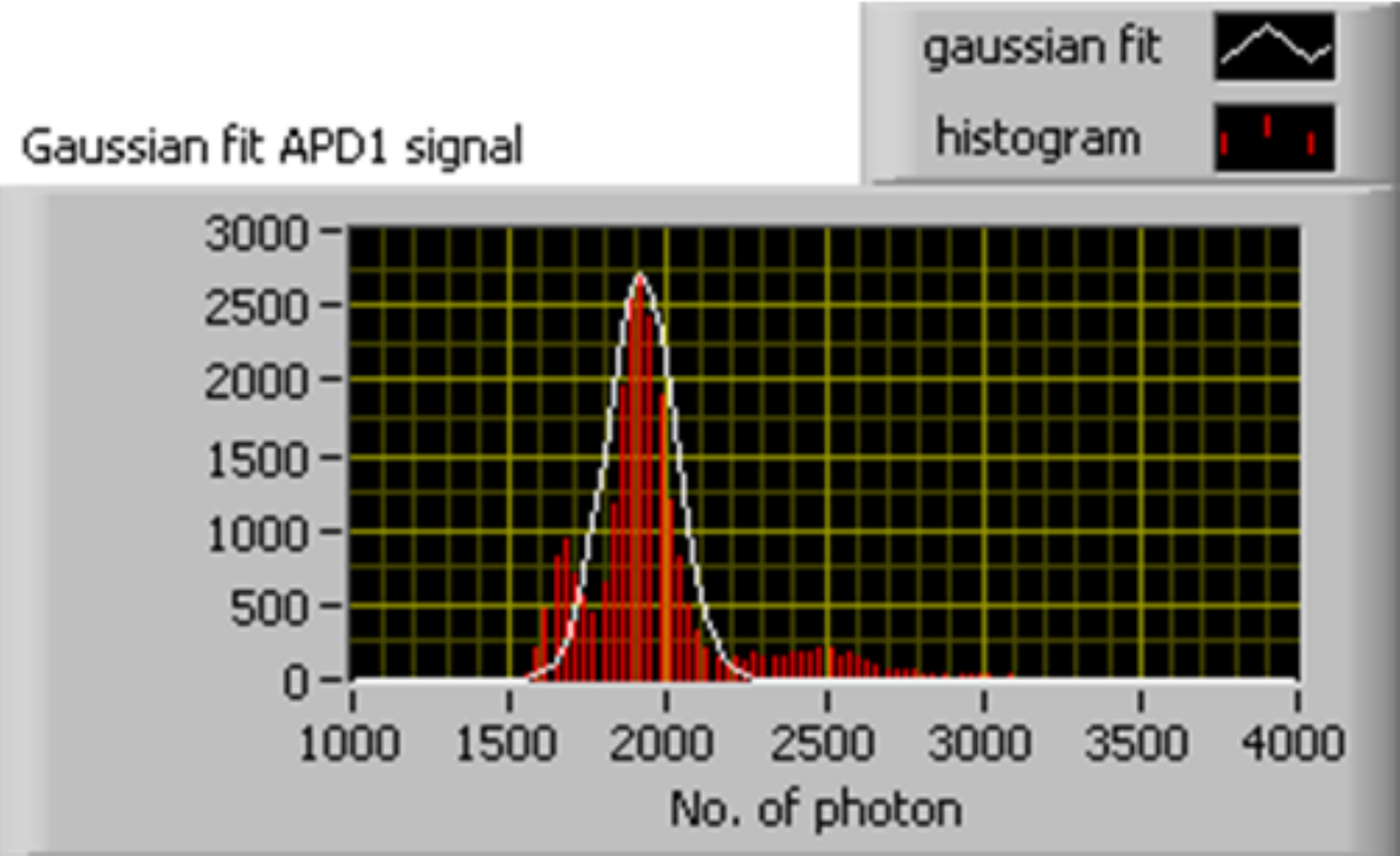}\label{apd11}}\hskip0.2cm\subfigure[]{\includegraphics[scale=0.2]{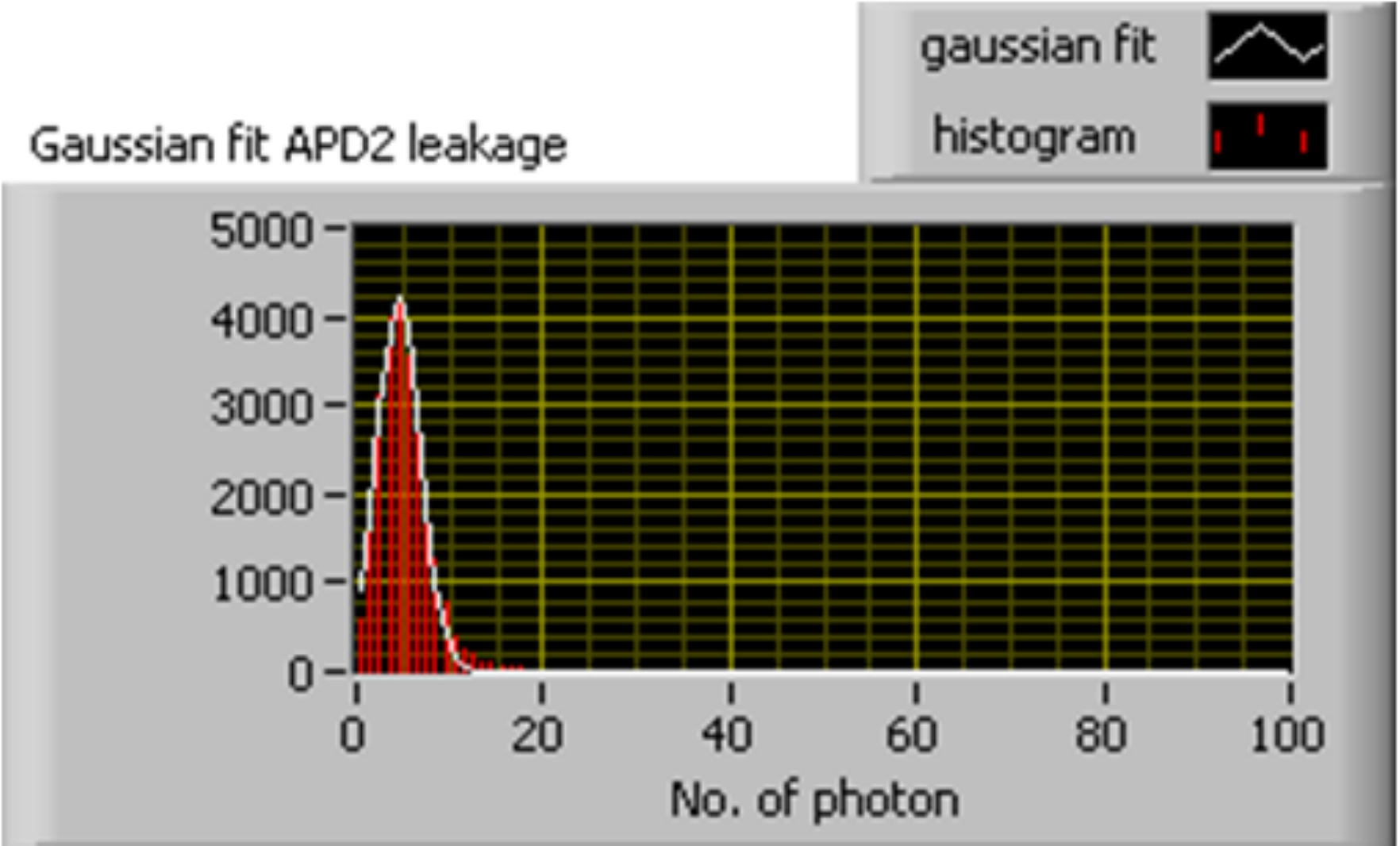}\label{apd12}}\\
\subfigure[]{\includegraphics[scale=0.2]{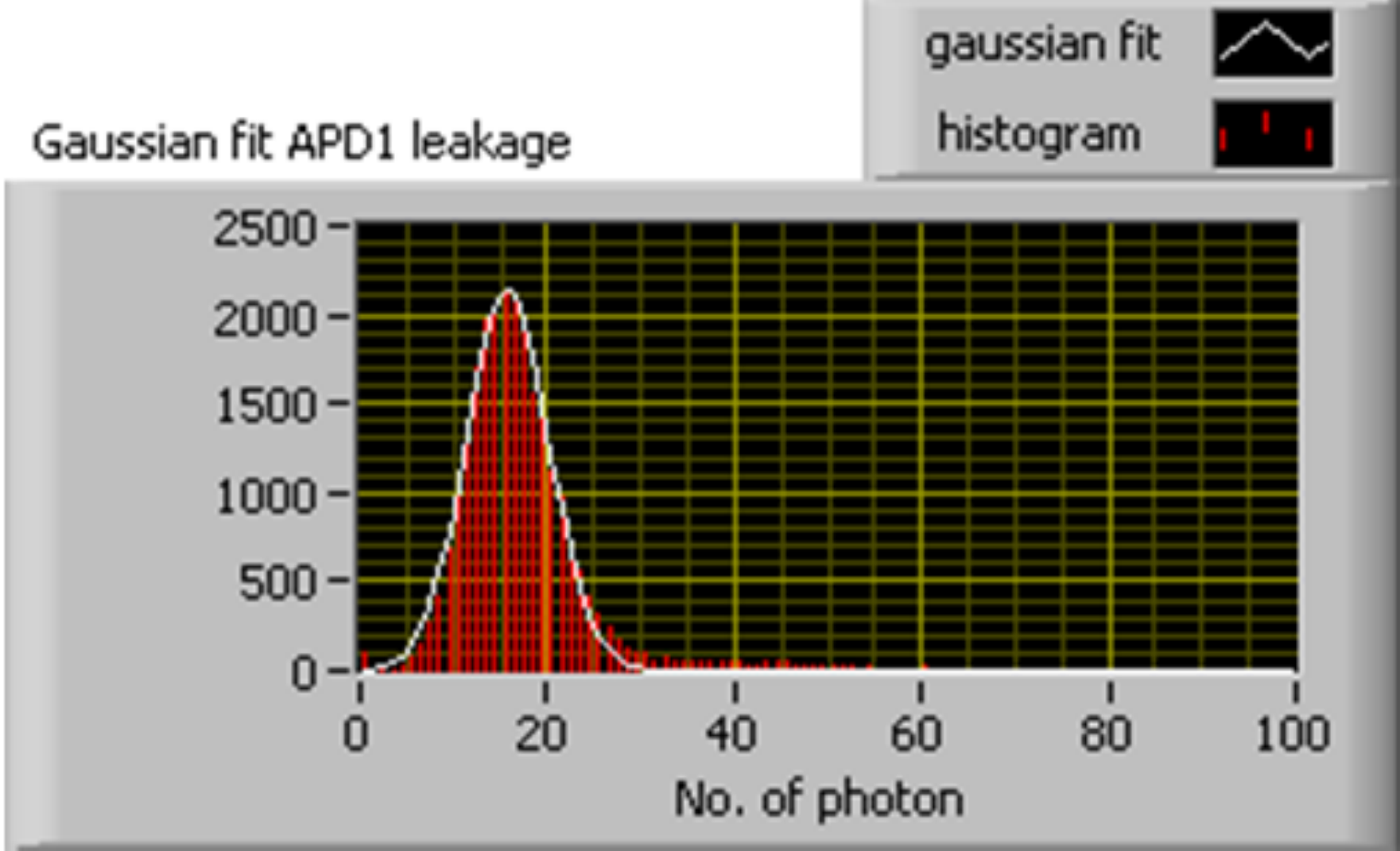}\label{apd21}}\hskip0.2cm\subfigure[]{\includegraphics[scale=0.2]{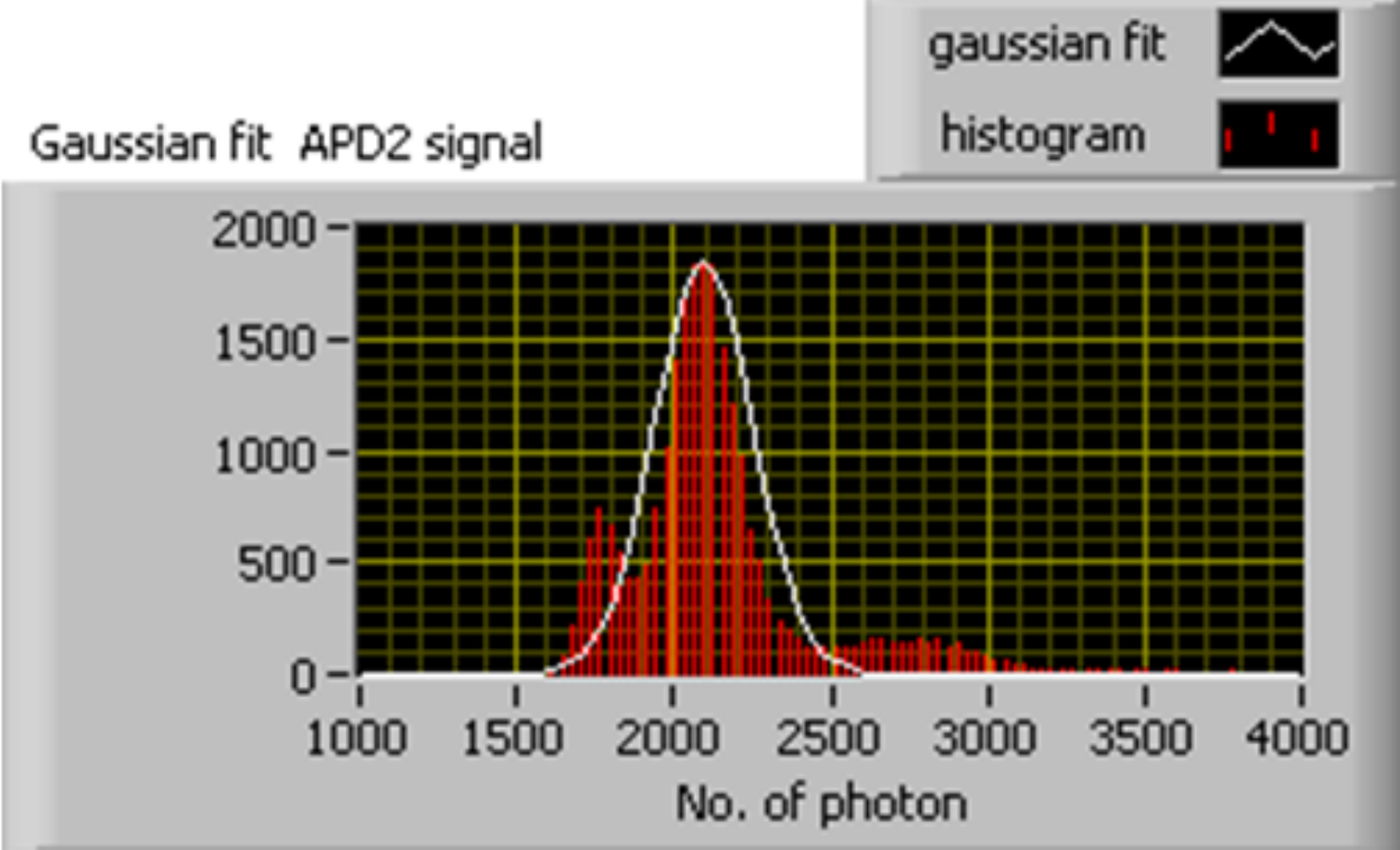}\label{apd22}}
\caption{Histograms of photon counts by APDs. (a) is true counts when L1 is on and  (b) is corresponding leakage. Similarly (c) is corresponding leakage when only laser L2 is on.}
\label{leakage}
\end{figure}

The experiment was performed by transferring about 1,00,000 bits of random sequence of 0's and 1's, first in Vertical/Horizontal basis (also called a `+' basis) and later in 45\degree/135\degree (`X' basis). Individual photon counts from APD1 and APD2 are recorded and the State of Polarization, SoP was computed as per formula \ref{sop}. The information bit was taken to be zero if SoP value was negative and bit was considered 1 if SoP was positive. By transmitting only one of the polarizations, we could compute the amount of leakage, obtain a histogram of the counts for true signal as well as leakage (see figure \ref{leakage}). 

. 

By fitting the statistical distribution of these data to individual Gaussian function, we obtain mean photon number for signal and noise/leakage as well as width of distribution. Using this in the standard formula for Quality factor  \cite{vorontsov, john}, as

\begin{equation}
Q=\frac{I_1-I_0}{\sigma_1+\sigma_0},
\label{q_factor}
\end{equation}

As in case of ON-off keying,  $I_1$ and $I_0$ stand for mean counts of signal and noise, given by centres of Gaussian distribution.   $\sigma_1$ and $\sigma_0$ are the width of respective Gaussian distributions.  But in this case there will be independent Q factors, one each for pair  polarizations. 


\begin{equation}
Q_i=\frac{I_i-I_j}{\sigma_i+\sigma_j}~~~~Q_j=\frac{I_j-I_i}{\sigma_j+\sigma_i}
\end{equation}

where $i,j$ are respectively $V,H$ or 45\degree/135\degree. 

The overall Q factor would be average of the individual Q factors above. The theoretical estimate of Bit error rates, as given by \cite{john}, for all possible $i$
\begin{equation}
{\rm BER_i} = \frac{{\rm erfc}(Q_i\sqrt{2})}{2}.
\label{ber}
\end{equation}

\subsection{The Quality factor}
\begin{figure}[!h]
\centerline{\includegraphics[scale=0.36,angle=270]{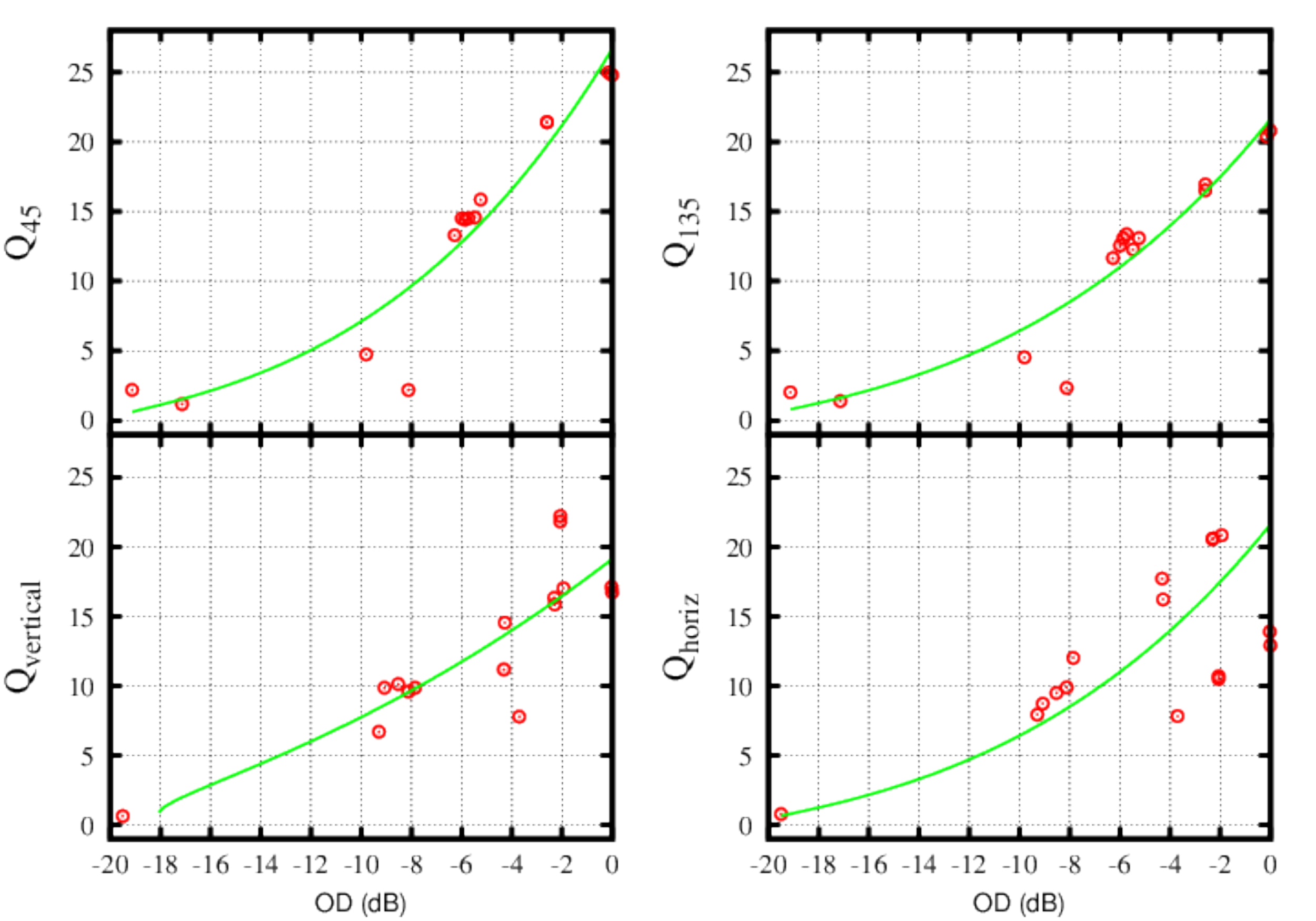}}
\caption{Variation of Q factors with respect to OD for smoke, vertical/horizontal and 45/135 degrees}
\label{smoke_Q_all}
\end{figure}

\begin{figure}[!h]
\centerline{\includegraphics[scale=0.36,angle=270]{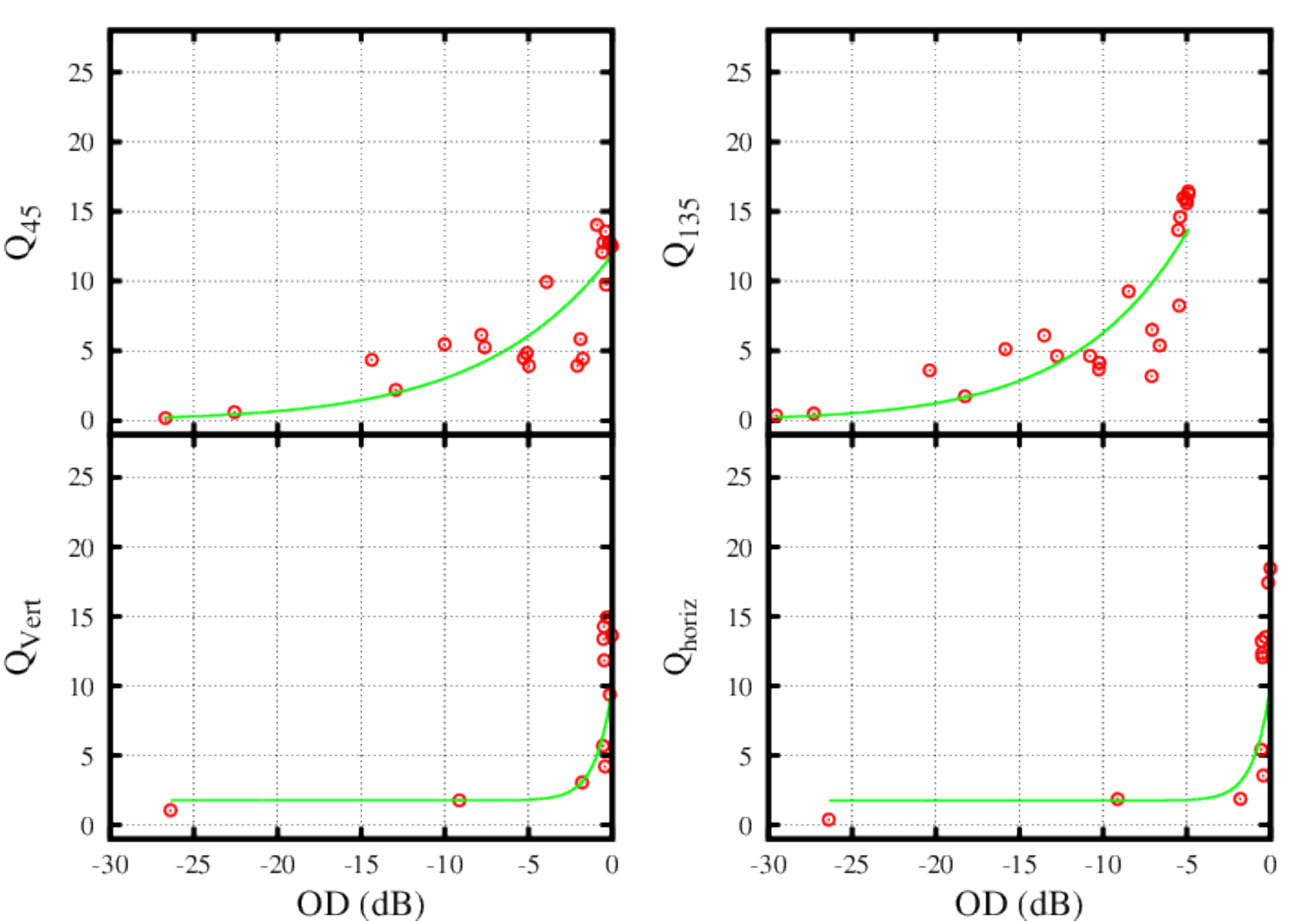}}
\caption{Variation of Q factors with respect to OD for fog, vertical/horizontal and 45/135 degrees}
\label{fog_Q_all}
\end{figure}

For the `+' basis transmission, about 10,000 data bits in random sequence of zeros and ones were transmitted and their individual SOP values were recorded, as per equation \ref{sop}.  Their statistical distribution was recorded as a histogram and fit to a Gaussian. From the parameters of Gaussian, Q factors were computed and a theoretical BER obtained using the formula \ref{ber}. These Q factors, as a function of OD are shown in figures \ref{smoke_Q_all} and \ref{fog_Q_all}.

%

The above data was all fit to a to a stretched exponential function 
$$Q=A\exp((x-x_0)^\beta+c,$$. Here,  $A$ is the amplitude, $x$ is the Optical Density, $x_0$ is the shift in scaling and $\beta$ is the stretch coefficient. The dependency is normal exponential for smoke, whereas it is a stretched exponential with stretch coefficient of $\beta \approx 0.4$ for vertical and horizontal polarizations, that too in case of smoke. The variation is not surprising since smoke and fog particles have different distributions in terms of size and scattering cross sections. Since the OD is a quantitative measure of number of scattering events per unit length, a higher OD corresponds to more scattering and therefore a stronger depolarization. As seen from the graphs, Q factor decreases as OD increases and asymptotically reaches zero. However, The exact physics behind this behaviour is needed to be explored further. Part of the answer may lie in the calculations of Moeyaert et. al. \cite{moeyaert}, who have shown that the BER in case of an OOK has a stretched exponential dependency.

In  case of 45\degree/ 135\degree, the  variation of Q factor with respect to OD is slower compared to the `H/V' basis, with both polarization modes showing a stretched exponential dependency with $Q=A\exp[-(x-x_0)^{0.5}$, where $x$ is the optical density with a shift of $x_0$. This is very critical in Quaternary encoding, since the 45\degree and 135 \degree polarized are depolarized more than the vertical/horizontal polarizations. However, it is evident that the differential method of computing SOP and then mapping to information bits has a higher tolerance for depolarization.
\subsection*{Bit Error Rate(BER)}

Theoretical estimate of BER, based on Q factor is given by equation \ref{ber}, which is more valid for a OOK scheme \cite{john}. However in this case we can define two BER values, one for each polarization - 

\begin{figure}[!h]
\subfigure[]{\includegraphics[scale=0.33]{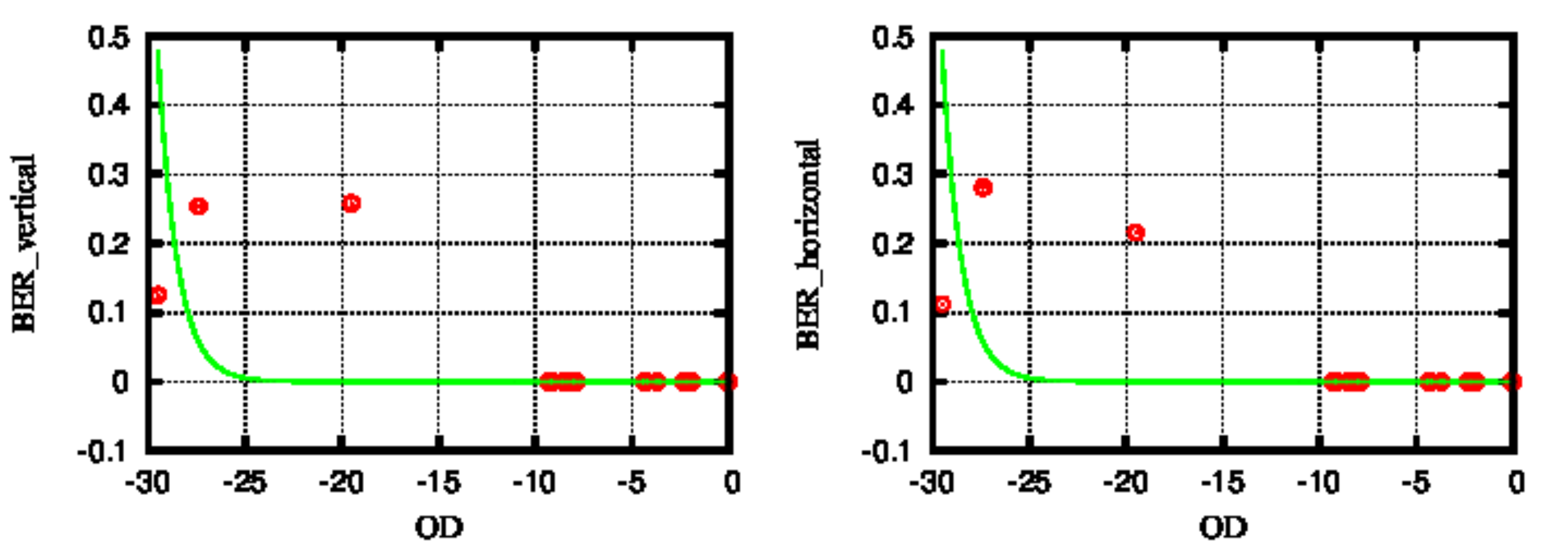}}
\subfigure[]{\includegraphics[scale=0.33]{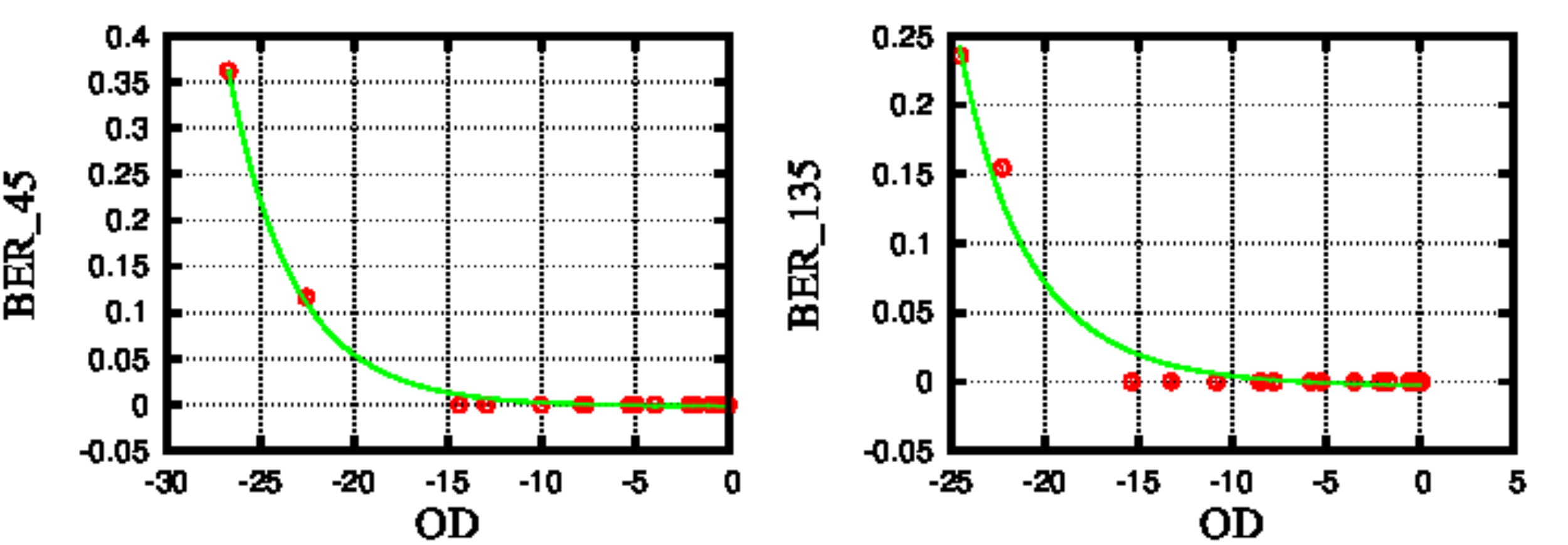}}
\caption{Estimated Bit Error rates in presence of smoke. Solid line indicates an exponential fit with stretch factor.  Values are reasonably low for almost upto OD=-20 dB.}
\label{BER_smoke}
\end{figure}

\begin{figure}[h]
\includegraphics[scale=0.33]{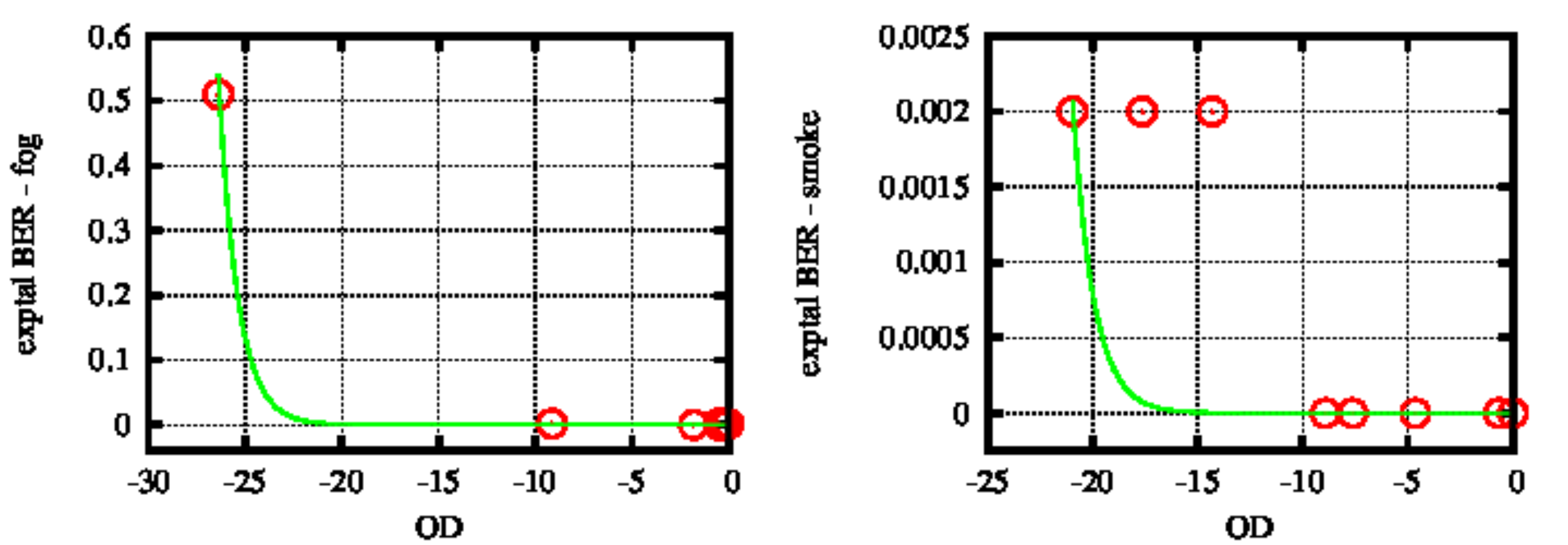}
\caption{Experimentally obtained BER from comparing transmitted and received data in presence of fog (left) and smoke (right). Solid lines are fit for exponential curve.}
\label{exp_ber}
\end{figure}

The estimated BER values are very close to zero for OD almost upto -20 dB, after which it raises sharply. The actual values of BER, which are obtained by comparing transmitted and received bits are shown in \ref{exp_ber}. These indicate a very similar trend, though the error rates for smoke are much lower than that for fog. This is due to the fact that fog causes a significantly higher depolarization than smoke. 

These results can be used to estimate a threshold for noise tolerance in using real communication situation. Instead of measuring actual bit error rates in real time, one can rely on the signal attenuation and propose that if the signal strength is less than -15 dB, the communication channel is likely to be noisy. An attenuation of more than 20 dB will definitely be unreliable.

\begin{figure}[!h]
\subfigure[]{\includegraphics[scale=0.35]{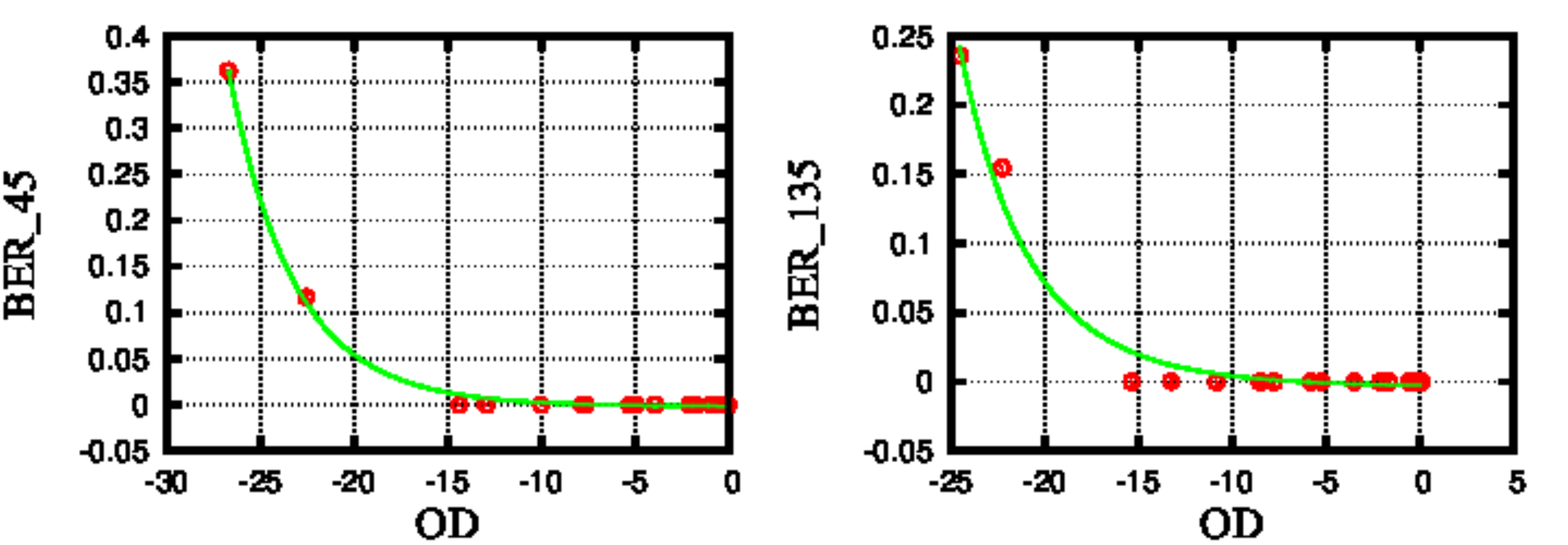}}
\subfigure[]{\includegraphics[scale=0.35]{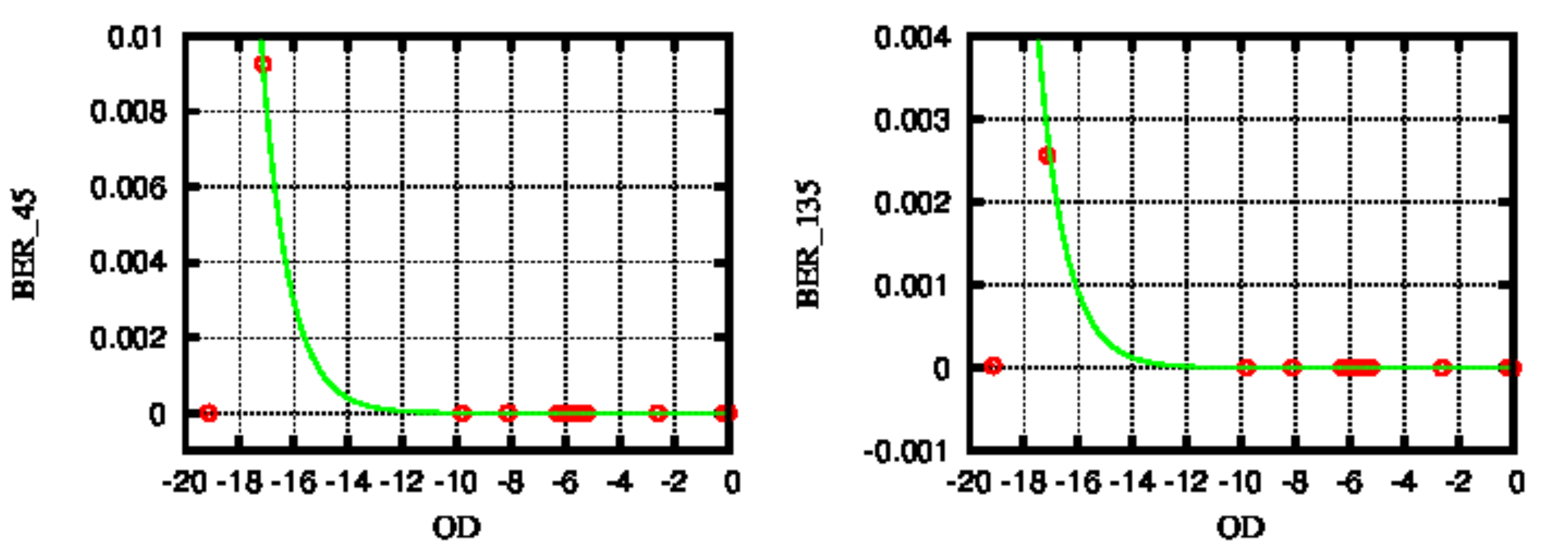}}
\caption{BER for 45$^\circ$ (left) and 135$^\circ$ (right), as a function of signal attenuation, in presence of (a) fog and (b) Smoke.  Solid line is fit for exponential function. As in case of vertical/horizontal case, BER is close to zero till a threshold value  of OD, and increases sharply beyond that. But this increase is much sharper and higher for smoke increases much more than in case of fog.}
\label{ber_45}
\end{figure}


 It was shown in reference \cite{soorat} that the value of SOP is determined only by snake and ballistic photons which retain their polarization and the depolarized part of the light will cancel out and does not affect the SOP. We show below the same results using a full Stokes vector analysis.

\section{Stokes Vector analysis}
Stokes vector picture gives a complete analysis for this situation since the depolarization due to atmosphere is properly represented by the corresponding M\"uller matrix. Stokes vector then gives a correct representation of a partially polarized light.  

We obtain the M\"uller matrices for all the optical components used, as $M_1$ for the first PBS, $M_2$ for the depolarizing field in free space  and the $M_3$ for the second PBS. Since PBS has two outputs, the corresponding M\"uller matrices are represented by   $M_i$ and $M_i^\prime$, with the primed components standing for horizontal and unprimed components indicating outputs at vertical port of PBS. They are given as 

$$
M_{1,3}=\frac{1}{2}\begin{pmatrix}
1&1&0&0 \\
1&1&0&0 \\
0&0&0&0 \\
0&0&0&0 \\ 
\end{pmatrix}, ~ M_2=\begin{pmatrix}
1&0&0&0 \\
0&a&0&0 \\
0&0&a&0 \\
0&0&0&a \\ 
\end{pmatrix}.
$$

and 

$$
M_1^\prime=M_3^\prime=\frac{1}{2}\begin{pmatrix}
1&-1&0&0 \\
-1&1&0&0 \\
0&0&0&0 \\
0&0&0&0 \\
\end{pmatrix},
$$

where, $a$ is the depolarization factor. For an ideal case of no depolarization, $a=1$ and $a=0$ for a complete depolarization.  In this situation, the outputs of  polarizing beam splitter ($M_3$ and $M_3^\prime$) are equivalent to taking projections of the input Stokes vector onto two different states, each of which gives a probability  of $0.5$. 

Net measurement from detectors $D_1$ and $D_2$ would then be

\begin{eqnarray*}
D_1=(M_3.M_2.M_1)L_1&+&(M_3.M_2.M_1^\prime)L_2 \cr \cr
D_2=(M_3^\prime.M_2.M_1)L_1&+&(M_3^\prime.M_2.M_1^\prime)L_2 
\end{eqnarray*}

$L_1\equiv\{1, 1, 0, 0 \}$ and $L_2\equiv\{1, -1, 0, 0 \}$ are the Stokes vectors for light from lasers L1 and L2.
During the  communication either L1 is fired or L2 is fired. The two lasers are never simultaneously on. Hence the stokes vector of the light at the detectors reduces to

\begin{eqnarray}
D_1&=&(M_3.M_2.M_1)L_1 \cr
D_2&=&(M_3^\prime.M_2.M_1)L_1 
\label{L1_on}
\end{eqnarray}

or 

\begin{eqnarray}
D_1&=&(M_3.M_2.M_1^\prime)L_2 \cr
D_2&=&(M_3^\prime.M_2.M_1^\prime)L_2
\label{L2_on}
\end{eqnarray}

Depending upon whether L1 or L2 is pulsed. Photodiodes measure total intensity of light, which corresponds to $S_0$ of the Stokes Vector. It can be shown from from equations \ref{L1_on} and \ref{L2_on}, that

%
%
%

for L1 case
\begin{equation}
D_1=\frac{1}{2}\begin{pmatrix} 1+a \\ 1+a \\ 0 \\ 0 \end{pmatrix} 
D_2=
\begin{pmatrix}
1-a\\
-1+a\\
0\\
0\\
\end{pmatrix} \\
\end{equation}

and for L2 case 

\begin{equation}
D_1=
\begin{pmatrix}
1-a\\
-1-a\\
0\\
0\\
\end{pmatrix} \\
D_2=
\begin{pmatrix}
1+a\\
-1-a\\
0\\
0\\
\end{pmatrix} \\
\end{equation}

Using this in \ref{sop}, for only the $S_0$ component of the Stokes vector, we get for L1 case as

\begin{equation}
\frac{\{D_1\}_{1}-\{D_2\}_{1}}{\{D_1\}_{1}+\{D_2\}_{1}} = \frac{1+a-1+a}{1+a+1-a} = a
\label{stokes_sop1}
\end{equation}

and the L1 case as

\begin{equation}
\frac{\{D_2\}_{1}-\{D_1\}_{1}}{\{D_2\}_{1}+\{D_1\}_{1}} = \frac{1-a-1-a}{1-a+1+a} = -a
\label{stokes_sop2}
\end{equation}

It is evident from above that the value of SOP is either  $-a$ or $+a$. With no depolarization this will be $\pm 1$, and reduces to  lower values when a depolarization occurs. If the light is completely depolarized, then $a=0$ and SOP can not be determined. 

Similarly, for the case of 45\degree and 135 \degree case, we can add the M\"uller matrix for half wave plate as well 
$$M_{4}=\frac{1}{2}\begin{pmatrix}
1&0&0&0 \\
0&0&1&0 \\
0&-1&0&0 \\
0&0&0&1 \\ 
\end{pmatrix}$$

The Stokes vector on the detector would therefore be 

\begin{eqnarray}
D_1&=&(M_3.M_2.M_4)L_1 \cr \cr
D_2&=&(M_3^\prime.M_2.M_4)L_1
\label{stokes4}
\end{eqnarray}

\begin{eqnarray}
D_1&=&(M_3.M_2.M_4)L_2 \cr \cr
{\rm and }&&\cr \cr
D_2&=&+(M_3^\prime.M_2.M_4)L_2
\label{stokes5}
\end{eqnarray}

respectively. The Corresponding Stokes vectors, now in 45\degree and 135\degree polarized light would be $L_1\equiv\{1, 0, 1, 0 \}$  and the $L_2\equiv\{1, 0, -1, 0 \}$. It can be shown that the final Stokes vector for light falling on the detectors would be 

\begin{equation}
D_1=\frac{1}{2}\begin{pmatrix} 1+a \\ 1+a \\ 0 \\ 0 \end{pmatrix}
D_2=\frac{1}{2} \begin{pmatrix} 1-a \\ -1+a \\ 0 \\ 0 \end{pmatrix}
\end{equation}

This gives an SOP as 
\begin{equation}
\frac{\{D_1\}_{1}-\{D_2\}_{1}}{\{D_1\}_{1}+\{D_2\}_{1}} = \frac{1+a-1+a}{1+a+1-a} = a
\label{stokes_sop11}
\end{equation}
Similarly when laser $L_2$ is on, we get 
\begin{equation}
\frac{\{D_1\}_{1}-\{D_2\}_{1}}{\{D_1\}_{1}+\{D_2\}_{1}} = \frac{-1+a+1+a}{-1+a-1-a} = -a
\label{stokes_sop21}
\end{equation}

As in case of vertical polarized light, the SOP values are either $-a$ or $+a$, which can be mapped to digits zero and one. This mapping is reliable as long as $a$ is not equal to zero. 

\section{Conclusion}
We have proposed a Quaternary encoding using four states of polarization, for use of free space Optical communication. To understand the nature of depolarization due to atmospheric effects such as fog and smoke, we created these effects within laboratory situation and studied it. It shows that behaviour of depolarization is different for fog and smoke, due to the difference in their constituent nature, but also for vertical and 45\degree basis. This can cause serious errors in a communication protocol. 

However, the Quality factors and Bit error rates obtained from a statistical distribution of the states of polarization, show a reasonably high Quality factor and a low BER, even for a significantly high scattering densities. In addition, a differential measurement method to map the final bit to the polarization allows even lower noise rates. The results shown will be very important in a practical implementation of a free space Quaternary encoding communication setup, which will be our plan for further studies. 

\section{Acknowledgement}
We thank Department of Information Technology, Govt. of India for financial help. RS thanks University Grants Commission of India for fellowship under RGNF scheme.

\end{document}